\documentclass[aps,showpacs,twocolumn]{revtex4}
\usepackage{epsfig}
\usepackage{color}
\usepackage{amsmath}
\usepackage{amssymb}
\usepackage{bbm}

\newcommand{\be}{\begin{equation}}
\newcommand{\ee}{\end{equation}}
\newcommand{\bea}{\begin{eqnarray}}
\newcommand{\beaa}{\begin{eqnarray*}}
\newcommand{\eea}{\end{eqnarray}}
\newcommand{\eeaa}{\end{eqnarray*}}

\begin{document}

\title{\bf\Large {Effect of Jahn-Teller coupling on Curie temperature in the Double Exchange Model}}

\author{Vasil Michev and Naoum Karchev\cite{byline}}

\affiliation{Department of Physics, University of Sofia, 1164 Sofia, Bulgaria}

\begin{abstract}
We consider the two-band double exchange model for manganites with Jahn-Teller (JT) coupling and explore the suppression of the ferromagnetism
because of the JT distortion. The localized spins of the $\emph{t}_{2g}$ electrons are represented in terms of the
Schwinger bosons, and two spin-singlet Fermion operators are
introduced instead of the $e_{g}$ electrons' operators.
In terms of the new Fermi fields the on-site Hund's interaction is in a diagonal form and one accounts for it exactly.
Integrating out the spin-singlet fermions, we derive an
effective Heisenberg  model for a vector which describes the
local orientations of the total magnetization.
The exchange constants are different for different space directions and depend on the density $n$ of $\emph{e}_{g}$ electrons and JT energy.
At zero temperature, with increasing the density of the $\emph{e}_{g}$ electrons the system undergoes phase transition from ferromagnetic phase $(0<n<n_c)$ to A-type antiferromagnetic phase $(n_c<n)$. The critical value $n_c$ decreases as JT energy is increased. At finite temperature we calculate the Curie temperature as a function of electron density for different JT energy. The results show that JT coupling strongly suppresses the spin fluctuations and decreases the Curie temperature.

\end{abstract}


\maketitle

Jahn-Teller (JT) effect is related to systems with degenerated electronic states \cite{KK}.
The importance of the JT coupling for manganites was first discussed in \cite{Millis1}
with regard to the colossal magnetoresistance. The most widely studied representatives have chemical formula $Re_{1-x}A_{x}MnO_3$, where $Re$ is rare earth such as $La$ or $Nd$, and $A$ is a divalent alkali such as $Ca$ or $Sr$. The important electrons in these compounds are $Mn$ $d$ electrons. They have five
degenerate levels \cite{Dagotto}. The crystal environment results in a particular splitting of the five
$d$-orbitals  (\emph{crystal field spliting}) into two groups: the
$\emph{e}_g$ and $\emph{t}_{2g}$ states. The electrons from the
$\emph{e}_g$ sector form a doublet, while the $\emph{t}_{2g}$ electrons form a triplet. The population of the $\emph{t}_{2g}$ electrons  remains constant, and
the Hund rule enforces alignment of the three $\emph{t}_{2g}$ spins into a
$S=3/2$ state. Then, the $\emph{t}_{2g}$ sector can be replaced by a
\emph{localized spin} at each manganese ion, reducing the complexity
of the original five orbital model. The only important interaction
between the two sectors is the Hund coupling between localized
$\emph{t}_{2g}$ spins and mobile $\emph{e}_g$ electrons.
The oxygens surrounding the manganese ion readjust their locations creating an asymmetry between the different directions. This effectively removes the degeneracy of the $e_{g}$ electrons' states. The lifting of the degeneracy due to the orbital-lattice interaction is called  Jahn-Teller effect.

The interaction between the electrons and phonons is unusually strong and leads to a wide range of striking physical phenomena. Changing the $\emph{e}_{g}$ electrons' concentration produces a variety of phases, which may be characterized by their magnetic, transport and charge-ordering properties \cite{CMR-Exp}. The manganites $La_{1-x}Ca_xMnO_3$ have attracted interest due to their colossal magnetoresistance. The phase boundary between ferromagnetism and paramagnetism, in these materials, also separates a low temperature metallic phase from a high-temperature insulating phase. At temperatures below Curie temperature $T<T_C$ the resistivity is relatively low and increases as T is increased, whereas at $T>T_C$ the resistivity is very high and (for most $x$) decreases as T is increased. The magnetoresistance for $T \approx T_C$ can be very large \cite{CMR-Exp}.

The double exchange model with JT coupling is a widely used model for manganites. The
procedures followed to obtain the essential features of the model are different: numerical studies \cite{Yunoki,Hotta}, Dynamical
Mean-Field Theory (DMFT) \cite{Millis2,Millis3,Held}, \emph{ab initio} density-functional calculations \cite{Popovic}, and analytical calculations \cite{Millis2,Millis3,Nolting1,Nolting2}.
In spite of the common conclusion that JT coupling suppresses the ferromagnetic state, the results are quite different and do not match the experimental results. For example the calculated Curie temperatures are two and even three times larger then the experimentally measured. Because of that it is important to formulate theoretical criteria for adequacy of the method of calculation. In our opinion the calculations should be in accordance with the Mermin-Wagner theorem \cite{M-W}. It claims that in two dimensions there is no spontaneous magnetization at nonzero temperature. Hence, the critical temperature should be equal to zero. We employ a technique of calculation \cite{M-K}, which captures the essentials of the magnon fluctuations in the theory, and for $2D$ systems one obtains zero Curie temperature, in accordance with Mermin-Wagner theorem. The physics of the ferromagnetic manganites near the Curie temperature is dominated by the magnon fluctuations and it is important to account for them in the best way.

The present paper is focused on the influence of the JT distortion on the ferromagnetism of manganites.
To model the manganites we employ the Hamiltonian
$H = H_{DE}  + H_{el-ph}$.
The first term describes the hopping of $e_g$ electrons and the Hund interaction between the spin $\textbf{s}_i$ of the $e_g$ electron and the localized $t_{2g}$ spin $\textbf{S}_i$
\be\label{JT2}
H_{DE} = \sum\limits_{i\, \textbf{a}\,ll'\alpha } {t_{ll'}^{\textbf{a}} c_{i l\alpha }^ +  c_{i+\textbf{a}\, l'\alpha } }- 2J_H \sum\limits_i {\textbf{s}_i \cdot \textbf{S}_i } \ee
where $c_{i l\alpha }^+ $ and $c_{i l\alpha }$ are creation and annihilation operators for $e_g$ electron with spin $\alpha$ in  $d_{x^2-y^2}(d_{3z-r^2}$) orbital at site $i$, and $\textbf{a}$ is the vector connecting nearest -neighbor sites. For the cubic lattice, the hopping amplitudes between $l$ and $l'$ orbitals along the $x,y,z$ directions are: \bea
t_{aa}^x  &=&  - \sqrt 3 t_{ab}^x  =  - \sqrt 3 t_{ba}^x  = 3t_{bb}^x  = t\nonumber\\
t_{aa}^y  &=& \sqrt 3 t_{ab}^y  = \sqrt 3 t_{ba}^y  = 3t_{bb}^y  = t\\
t_{aa}^z  &=& t_{ab}^z  = t_{ba}^z  = 0,\quad t_{bb}^z  = 4t/3\nonumber
\eea
The second term in Eq.\eqref{JT2} is the Hund interaction between the spin $\textbf{s}_i$ of the $e_g$ electron and the localized $t_{2g}$ spin $\textbf{S}_i$
with $s^{\nu}_i = 1/2\sum\limits _{l\alpha\beta} c_{i l\alpha }^ +\sigma^{\nu}_{\alpha\beta} c_{il\beta}$, where $\sigma^x,\sigma^y,\sigma^z$ are Pauli matrices, and the Hund's constant is positive $(J_H>0)$.

The $H_{el-ph}$ Hamiltonian models the coupling of $e_g$ electrons to the lattice distortion
\be\label{JT3}
\hskip -.2cm H_{el-ph}  = g\sum\limits_i \left(Q_{2i} \tau _{xi}  + Q_{3i} \tau _{zi} \right)+ \frac{k}{2}\sum\limits_i {\left( Q_{2i}^2  + Q_{3i}^2 \right)}\ee
where
$\tau _{xi}  = \sum\limits_{\alpha}  \left( c_{ia\alpha }^ +  c_{ib\alpha }  + c_{ib\alpha }^ +  c_{ia\alpha } \right)$ and $\tau _{zi}  = \sum\limits_{\alpha} \left(c_{ia\alpha }^ +  c_{ia\alpha}  - c_{ib\alpha}^ +  c_{ib\alpha}\right)$. In equation \eqref{JT3} $g$ is the electron-phonon coupling constant, while  $Q_{2i}$ and $Q_{3i}$ are JT phonon modes.
The second term in $H_{el-ph}$ is the usual quadratic potential for distortions with constant $k$. The important energy scale of the phonon-electron interaction is the static JT energy $E_{JT}=g^2/(2k)$.

One can represent the spin operators $\textbf{S}_i$ of the localized $t_{2g}$ electrons in terms of Schwinger-bosons ($\varphi_{i\alpha},
\varphi_{i\alpha}^{\dagger}$)
$S^{\nu}_i  =
\frac{1}{2}\varphi _{i\alpha }^ +  \sigma^{\nu}_{\alpha \beta} \varphi
_{i\beta},\quad \varphi _{i\alpha }^ + \varphi
_{i\alpha}  = 2s$. By means of the Schwinger-bosons we introduce spin-singlet Fermi fields
\bea\label{Fsinglet} &&\hskip -1cm\Psi^A_{il}(\tau)=\frac {1}{\sqrt
{2s}}\varphi^+_{i\alpha}(\tau)c_{il\alpha}(\tau)\label{cfm8}\\
&& \hskip -1cm \Psi^B_{il}(\tau)=\frac {1}{\sqrt
{2s}}\left[\varphi_{i1}(\tau)c_{il2}(\tau)\,-
\,\varphi_{i2}(\tau)c_{il1}(\tau)\right] \label{cfm9} \eea
and write the spin of the $e_g$ electron and the total spin of the system $\textbf{S}^{\rm tot}_{i} =  \textbf{S}_i+\textbf{s}_{i}$ in terms of the singlet fermions \cite{M-K}. Further, we average  the total spin of the system in
the subspace of the singlet fermions $A$ and $B$. The vector $\textbf{M}_i= \langle \textbf{S}^{\rm
tot}_i \rangle _f$ identifies
the local orientation of the total magnetization. Because of the fact that $t_{2g}$-electron spin is parallel with $e_{g}$-electron spin we obtain
$\textbf{M}_i = \frac MS \textbf{S}_i$ with
$M = S+\frac 12 \sum\limits_l \langle \left
(\Psi^{A+}_{il}\Psi^A_{il}-\Psi^{B+}_{il}\Psi^B_{il}\right) \rangle_f $.
Now, if we use Holstein-Primakoff representation for the vectors
$\textbf{M}_i(a^+,a)$ with $M$ as an "effective spin" of the system
$(\textbf{M}_i^2=M^2)$,  the bose fields
$a_i$ and $a^+_i$ are the \textbf{true magnons} in the system.

An important advantage of working with singlet fermions is the fact
that in terms of these spin-singlet fields the spin-fermion
interaction is in a diagonal form, the spin variables (magnons) are
removed, and one accounts for it exactly.
The theory is quadratic with respect to the spin-singlet
fermions and one can integrate them out to obtain the free energy of fermions as a function of the magnons' fields $a_i^+, a_i$. We expand
the free energy in powers of magnons' fields and keep only the first two terms. The first term $\emph{F}_{f0}$, which does not depend on the magnons' fields, is a free energy of Fermions with spins of localized $t_{2g}$ electrons treated classically. We fix the model parameters and consider this term as a function of the JT distortion modes independent on the lattice sites . The numerical calculations shows that the function depends only on $\sqrt{Q_{2}^2+Q_{3}^2}$ and we set $Q_{3}=0$. The physical value of the JT distortion is the value at which $\emph{F}_{f0}$ has a minimum. In this way we obtain the distortion as a function of the density of $e_g$ electrons for different values of JT energy and fixed Hund's coupling. We fix the hopping parameter $t=1$ to set the energy unit. The results for the renormalized distortion $Q=gQ_2$ as a function of charge carrier density $n$ are plotted in Fig. 1, for different values of the JT energy $E_{JT}$ and $J_H=15$.
\begin{figure}[!ht]
\epsfxsize=\linewidth
\epsfbox{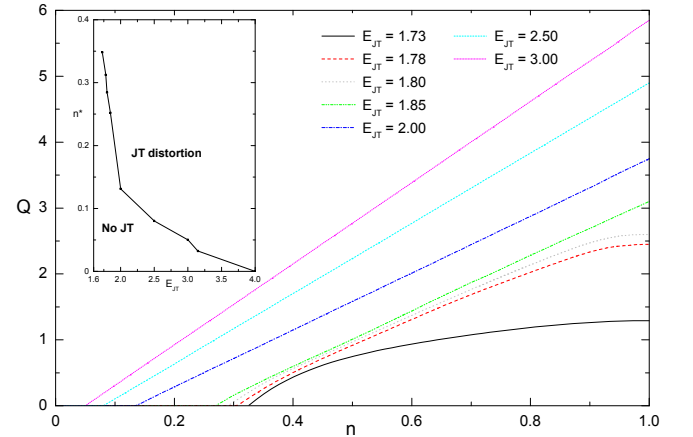} \caption{(Color online) The renormalized distortion $Q=gQ_2$ as a function of density $n$ for $J_H=15$ and differnt values of the JT energy $E_{JT}$. Inset: The density $n^*$, at which the JT distortion appears, as a function of JT energy $E_{JT}$. }\label{figQ}
\end{figure}
The figure \eqref{figQ} shows that JT distortion appears at critical value of the charge carrier density $n^*$ and increases as density $n$ is increased. The inset demonstrates that $n^*$ decreases and approaches zero as JT energy $E_{JT}$ is increased.

\begin{figure*}[!ht]
\epsfxsize=\linewidth 
\epsfbox{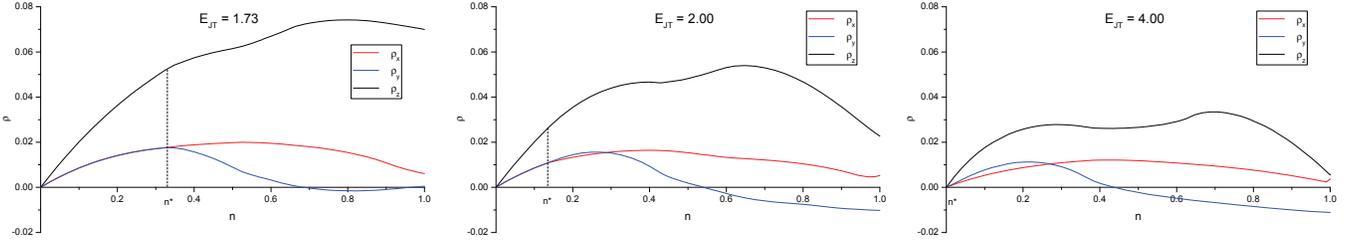} \caption{(Color online) Spin stiffness constants as a functions of density n for $J_{H}=15$, $E_{JT}=1.73$ (left), $E_{JT}=2$ (middle) and $E_{JT}=4$ (right). The vertical dash lines correspond to the density $n^*$ at which the JT distortion appears.}\label{fig2a}
\end{figure*}

The second term in the Fermion free energy is quadratic with respect to the magnons' fields $a_i^+, a_i$ and defines the effective magnon Hamiltonian in Gaussian approximation.
\be\label{JTeff} H_{\rm eff} = \sum\limits_{i \textbf{a}} \rho^{\textbf{a}}\left(a_i^+a_i + a_{i+\textbf{a}}^+a_{i+\textbf{a}} - a_i^+a_{i+\textbf{a}} - a_{i+\textbf{a}}^+a_i \right)\ee
In equation \eqref{JTeff} $\rho^{\textbf{a}}$ are spin stiffness constants which depend on the space directions $\textbf{a}$. They are calculated at zero temperature, for fixed Hund's coupling, JT energy, charge density, and JT distortion determined above. The calculations follow the technique developed in \cite{M-K}.
Based on the rotational symmetry, one can
supplement the Hamiltonian \eqref{JTeff} up to an effective Heisenberg like Hamiltonian, written in terms of the vectors
$\textbf{M}_i$
\be\label{JTeff2} H_{\rm eff}= - \sum\limits_{i \textbf{a}} J^{\textbf{a}}\textbf{M}_i\cdot \textbf{M}_{i+\textbf{a}} \ee where $J^{\textbf{a}}=\rho^{\textbf{a}} /M$. The
ferromagnetic phase is stable if all effective exchange coupling
constants are positive $J^{\textbf{a}} > 0$ $(\rho^{\textbf{a}}>0)$. If one of them is negative, for example $J^y<0\,\, (\rho^y<0)$, and the others are positive $J^x>0,\,J^z>0\,\, (\rho^x>0,\rho^z>0)$, the stable state is A-type antiferromagnetic phase which has planes $(x,z)$ that are ferromagnetic (parallel moments), with antiferromagnetic (antiparallel) moments between them. The spin-stiffness constant, as a function of charge carrier density, is depicted in Fig.\eqref{fig2a} for $J_H=15$, and three different values of JT energy, $E_{JT}=1.73,\,\,E_{JT}=2$ and $E_{JT}=4$. The vertical dash lines correspond to the density $n^*$ at which the JT distortion appears.
The figure on the left illustrates in the best way the impact of the JT distortion on the spin stiffness constants. The appearance of the distortion at $n^*$ is accompanied with a change of the slopes of the curves. The distortion splits the $\rho^y$ (blue) and $\rho^x$ (red) lines, and $\rho^y$ starts to decrease. At critical density $n_c$,  $\rho^y$ becomes equal to zero and the system undergoes a transition from ferromagnetic phase to A-type antiferromagnetic phase. The two other figures shows that spin stiffness constants decrease when JT energy increases and the critical density $n_c$ decreases too. As the spin stiffness constants are a measure for the magnon fluctuations in the ferromagnetic phase we conclude that JT distortion suppresses the magnon fluctuations.

The most evident consequence of this suppression is the Curie temperature ($T_C$) decreasing. To calculate $T_C$ we utilize the Schwinger-bosons
mean-field theory \cite{S-b1}.
We represent the vector $\textbf{M}_i$ Eq.\eqref{JTeff2} by
means of Schwinger bosons ($\phi_{i\alpha},\phi_{i\alpha}^+$)
\be
M^{\nu}_i = \frac{1}{2}\sum\limits_{\alpha \beta} {\phi _{i\alpha
}^+ \sigma^{\nu}_{\alpha \beta} } \phi _{i\beta}\qquad\phi
_{i\alpha}^+ \phi _{i\alpha} =2M\ee Next we use the identity
\be
\textbf{M}_i
\cdot \textbf{M}_j  = \frac{1}{2}\left( {\phi _{i\alpha}^ +  \phi
_{j\alpha} } \right)\left( {\phi _{j\beta}^ +
\phi _{i\beta} } \right)
-\frac{1}{4}\left( {\phi _{i\alpha}^ +  \phi _{i\alpha} }
\right)\left( {\phi _{j\beta}^ +  \phi _{j\beta} }\right)\ee
and rewrite the effective Hamiltonian in the form
\be\label{JTeff3} H_{\rm eff}  =  - \frac{1}{2}\sum\limits_{i\textbf{a} }J^{\textbf{a}} {\left( {\phi _{i\alpha
}^ +  \phi _{i+\textbf{a}\alpha} } \right)\left( {\phi _{i+\textbf{a}\beta}^ +  \phi
_{i\beta} } \right)}\ee where the constant term is dropped. To ensure the constraint
we introduce a parameter ($\lambda$) and add a new term
to the effective Hamiltonian \eqref{JTeff3}.
\be \label{JTeff4} \hat H
_{\rm eff} = H_{\rm eff}  + \lambda \sum\limits_i {\left( {\phi
_{i\sigma }^ + \phi _{i\sigma }  - 2M} \right)}\ee
We treat the four-boson interaction within Hartree-Fock
approximation. The Hartree-Fock hamiltonian which corresponds to the
effective hamiltonian reads \bea \label{JTH-F}
& & H_{\rm H - F}= \frac{1}{2}\sum\limits_{i\textbf{a}} J^{\textbf{a}}\bar u_{i,i+\textbf{a}}
u_{i,i+\textbf{a}} +\lambda \sum\limits_i {\left( {\phi _{i\sigma }^ +  \phi
_{i\sigma }  - 2M} \right)}\nonumber \\ & - & \frac{1}{2}\sum\limits_{i\textbf{a}} J^{\textbf{a}}{\left[ {\bar u_{i,i+\textbf{a}}
\phi _{i\alpha}^ +
\phi _{i+\textbf{a}\alpha}  + u_{i,i+\textbf{a}} \phi _{i+\textbf{a}\alpha}^ +  \phi _{i\alpha} } \right]}
 \eea where $\bar u_{i,i+\textbf{a}}\, ( u_{i,i+\textbf{a}})$ are
Hartree-Fock parameters to be determined self-consistently. We are
interested in real parameters which do not depend on the lattice
sites, but depend on the space directions $u_{i,i+\textbf{a}}=\bar u_{i,i+\textbf{a}}=u_{\textbf{a}}$. Then in momentum space
representation, the Hamilonian \eqref{JTH-F} has the form
\be\label{JTH-F2}
H_{\rm H - F}  = \frac{{N}}{2}\sum\limits_{\textbf{a}}u_{\textbf{a}}^2J^{\textbf{a}}  - 2\lambda MN+ \sum\limits_k {{\varepsilon _k}\phi^+_k\phi_k },\ee where $N$ is the number of lattice sites and $\varepsilon_k$  is the dispersion of the $\phi_k$-boson (spinon).
The free energy of the theory with Hamiltonian $H_{\rm H - F}$ is
\be\label{JTfreeE}
F = \frac 12 \sum\limits_{\textbf{a}}u_{\textbf{a}}^2J^{\textbf{a}}   - 2\lambda M + \frac{{2T}}{N} \sum\limits_k {\ln
\left( {1 - e^{ - \frac{{\varepsilon _k }}{T}} } \right)},\ee where $T$ is the temperature. The equations for the parameters
$u_{\textbf{a}}$ and $\lambda$ are:
${\partial F}/{\partial u_{\textbf{a}}}=0 \quad{\partial F}/{\partial \lambda}=0$.

To solve the system of four equations it is more convenient to introduce a new parameter ($\mu$) instead of ($\lambda$):
$\lambda  = \sum\limits_{\textbf{a}}\left(u_{\textbf{a}} J^{\textbf{a}}+ \mu u_{\textbf{a}}\right) $. In terms of the new parameter the $\phi_k$-boson dispersion is
$\varepsilon_k = \sum\limits_{\textbf{a}} \left[u_{\textbf{a}} J^{\textbf{a}} \left(1-\cos k_{\textbf{a}}\right)+\mu u_{\textbf{a}}\right]$ and the theory is well defined for positive constants $u_{\textbf{a}}\geq0$ and $\mu \geq 0$. For high enough temperatures
$\mu(T)$ and $u_{\textbf{a}}(T)$ are positive, and the excitation is gapped.
Decreasing the temperature leads to decrease of $\mu(T)$. At temperature $T_C$
it becomes equal to zero $\mu(T_C)=0$, and long-range excitation emerges in the
spectrum. Therefore this is the Curie temperature.
\begin{figure}[!t]
\epsfxsize=\linewidth
\epsfbox{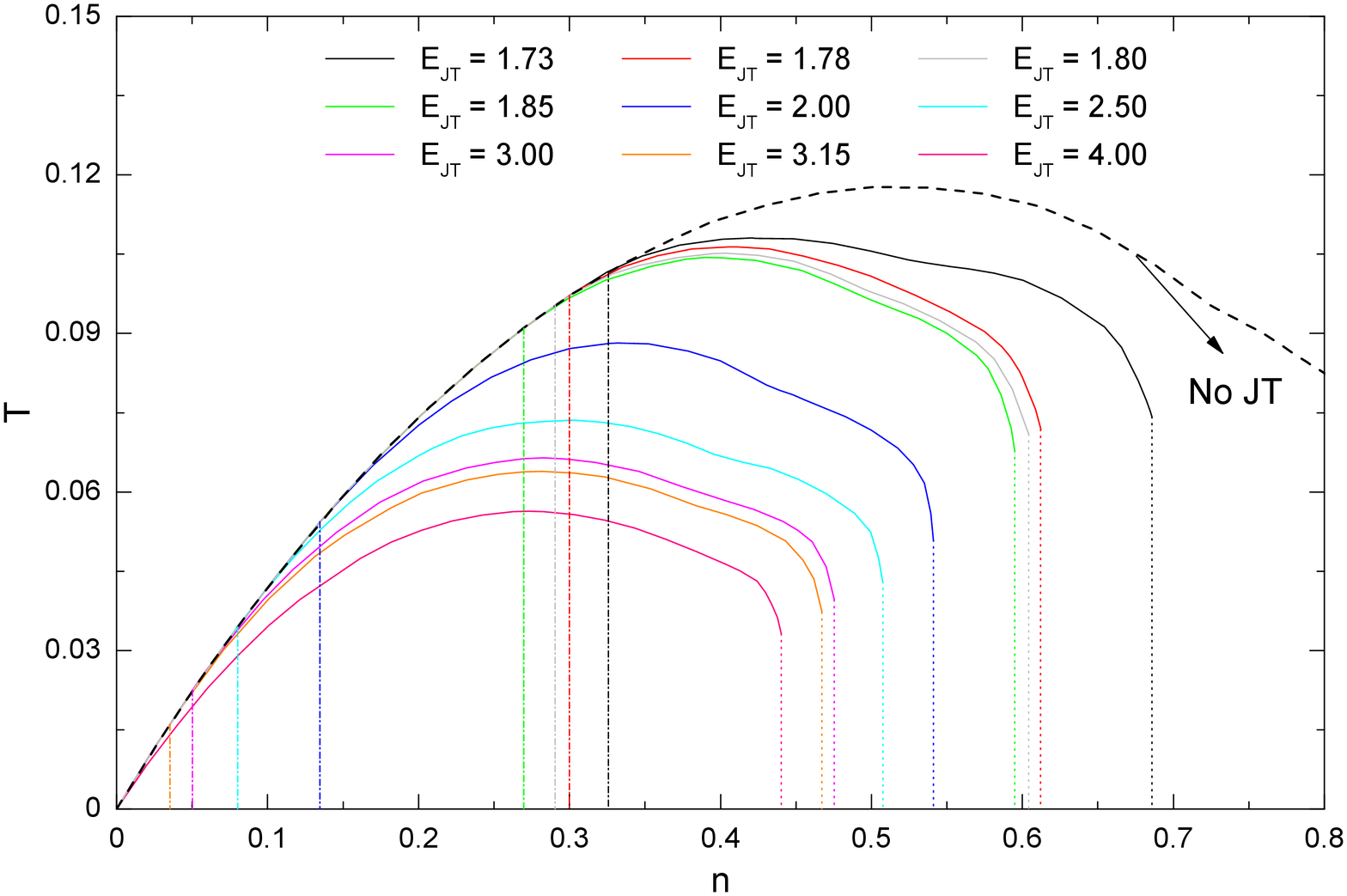} \caption{(Color online) $T_C$ as a function of $e_g$ electron density $n$ for $J_H=15$ and different values of the JT energy.}\label{JTTC}
\end{figure}
We set $\mu=0$ and obtain a system of equations for the
Curie temperature $T_C$ and $u_{\textbf{a}}$
\bea\label{JTCurie}
 & & u_{\textbf{a}'} = \frac{2}{N}\sum\limits_k \frac{\cos k_{\textbf{a}'}}{e^{\frac{1}{MT_C}\sum\limits_{\textbf{a}} u_{\textbf{a}} \rho^{\textbf{a}}(1-\cos k_{\textbf{a}})}  - 1} \\
 & & M = \frac{1}{N}\sum\limits_k \frac{1}{e^{\frac{1}{MT_C}\sum\limits_{\textbf{a}} u_{\textbf{a}} \rho^{\textbf{a}}(1-\cos k_{\textbf{a}})}  - 1} \nonumber\eea
 The results for the Curie temperature $T_C$ as a function of $e_g$ electrons density $n$ are plotted in figure \eqref{JTTC}, for $J_H=15$ and different values of the JT energy. The upper (black) dash line is a reference line which corresponds to the case without JT distortion. The vertical dash dot lines, on the left, correspond to the density $n^*$, while the vertical dot lines, on the right, correspond to the critical density $n_c$. The appearance of the JT distortion at $n^*$ leads to a spitting of the reference curve and the curve for a system with JT distortion. The density $n^*$ decreases when JT energy  increases and the ferromagnetic phase is strongly suppressed because of the suppression of the magnon fluctuations, which in turn leads to the decreasing of the Curie temperature.

To illustrate our results we present in a table the maximal Curie temperatures ($T^{max}_C[K]$), for different JT energies ($E_{JT}/t$) and the corresponding $e_g$-electron densities ($n$). To do this we have utilized that $t=0.8eV$ \cite{Held}.
\vskip 0.25cm
\begin{centering}\begin{tabular}{|c|c|c|c|c|c|c|c|c|c|}\hline
$E_{JT}/t$&1.73&1.78&1.80&1.85&2.00&2.50&3.00&3.15&4.00\\ \hline
$n$&0.42&0.41&0.40&0.39&0.33&0.30&0.283&0.282&0.27\\ \hline
$T_C^{\rm max}[K]$&1004&988&976&970&818&684&618&594&522\\ \hline
\end{tabular}\end{centering}
\vskip 0.25cm
We have used a large value for Hund's constant to better demonstrate the impact of the JT distortion on the ferromagnetism.
Decreasing of $J_H$ suppresses the ferromagnetic phase, decreases the Curie temperature, and reduces the impact of the JT distortion on the ferromagnetism. For example, for $J_H/t = 15$ and absence of JT distortion we have $T_C=1092K$, while for $J_H/t = 5$ we obtain $T_C=738K$. For non-zero distortion, $E_{JT}/t=2$, the Curie temperatures are $T_C=818K$ and $T_C=620K$ respectively. These results show that the reduction of the Curie temperature due to JT distortion depends on the value of $J_H$.

The authors acknowledge the financial support of the Sofia
University. This work was partly supported by a Grant-in-Aid DO02-264/18.12.08 from NSF-Bulgaria.

\end{document}